# Evaluation of Thermal Performance of a Wick-free Vapor Chamber in Power Electronics Cooling

Arani Mukhopadhyay [1], Anish Pal [1], Congbo Bao [2], Mohamad Jafari Gukeh [1], Sudip K. Mazumder [2], Constantine M. Megaridis [1,*]
[1] *Mechanical and Industrial Engineering, University of Illinois Chicago, IL, United States*
[2] *Electrical and Computer Engineering, University of Illinois Chicago, IL, United States*
* cmm@uic.edu

*Abstract*—Efficient thermal management in high-power electronics cooling can be achieved using phase-change heat transfer devices, such as vapor chambers. Traditional vapor chambers use wicks to transport condensate for efficient thermal exchange and to prevent "dry-out" of the evaporator. However, wicks in vapor chambers present significant design challenges arising out of large pressure drops across the wicking material, which slows down condensate transport rates and increases the chances for dry-out. Thicker wicks add to overall thermal resistance, while deterring the development of thinner devices by limiting the total thickness of the vapor chamber. Wickless vapor chambers eliminate the use of metal wicks entirely, by incorporating complementary wettability-patterned flat plates on both the evaporator and the condenser side. Such surface modifications enhance fluid transport on the evaporator side, while allowing the chambers to be virtually as thin as imaginable, thereby permitting design of thermally efficient thin electronic cooling devices. While wick-free vapor chambers have been studied and efficient design strategies have been suggested, we delve into real-life applications of wick-free vapor chambers in forced air cooling of high-power electronics. An experimental setup is developed wherein two Si-based MOSFETs of TO-247-3 packaging having high conduction resistance, are connected in parallel and switched at 100 kHz, to emulate high frequency power electronics operations. A rectangular copper wick-free vapor chamber spreads heat laterally over a surface 13 times larger than the heating area. This chamber is cooled externally by a fan that circulates air at room temperature. The present experimental setup extends our previous work on wick-free vapor chambers, while demonstrating the effectiveness of low-cost air cooling in vapor-chamber enhanced high-power electronics applications.

## I. Introduction

The advent of better performing wide bandgap semiconductor (WBG) devices that can switch at very high frequencies, often greater than 100kHz [1, 2], calls for the development of efficient thermal cooling and management strategies. Greater switching frequencies have cut down the physical dimensions of power electronics systems, thereby enabling smaller devices with faster response times and better control of device function. However, such reduction in size, also requires better dissipation of the generated heat, as performance of such WBG devices is greatly affected by device temperature. Smaller devices generate heat in extremely confined spaces (and hence, generate greater heat fluxes), and common methods of heat dissipation via single-phase cooling are not sufficient to maintain optimum device temperatures. Liquid cooling has been an impediment to design of high-power electronic packages with power densities greater than $10^7$ W/m$^2$, requiring efficient removal of heat from semiconductor junctions [3]. Furthermore, liquid cooling also entails complexities in the form of large device sizes, additional operational costs, and increased system weight and dimensions. Herein Vapor Chambers (VCs), which are passive heat-spreading devices that use two-phase heat transfer for effectively spreading heat over the length of the device, can be used for handling such heat fluxes. VCs are hermetically sealed devices that hold a fixed amount of a liquid at pressures lower than the atmosphere, and do not contain non-condensable gases (NCGs). The heat from the heat source (e.g., a semiconductor device) evaporates the liquid inside the vapor chamber, which then condenses on the other side, and flows down to the evaporator either via a wick [4] or via specially designed wettability-engineered patterns [5]. This process keeps on moving in cycles that help spread heat effectively in two-dimensions. Such VCs can be coupled with fins, or other cooling technology (viz. liquid-cooled plates) that can effectively remove the heat from the condenser. Air cooling holds significant advantages over liquid cooling, such that this method does not add complex parts to the system, and also does not entail issues related to leakage of the coolant.

Different designs of VCs have been studied extensively over the last few decades, many of which include modifications to the ways in the working fluid is transported to the evaporator. Recent studies like those by Koukoravas et al. [4] or by Gukeh et al. [6] in cooling of power electronic applications, have incorporated the use of wicks to return the working fluid to the evaporator, while maintaining a uniform thin layer over the hot side of the vapor chamber, so that the liquid can freely transition into the vapor space. Other approaches like those by Bandyopadhyay et al. [7], have incorporated the use of cascaded wicked multi-cores, that can sequentially transfer the heat from one vapor chamber to another, without the possibility of a thermal dry-out, even at high heat loads. Thinner vapor chambers, like those used by Li et al. [8], use supporting columns for maintaining the vapor chamber structure. Furthermore, Li et al. investigated the effect of the number of supporting columns to the heat transfer performance of their vapor chambers. Significant investigations have also focused into developing VCs with wickless components. Studies by Jafari Gukeh et al. [9], have demonstrated the applicability of wettability patterned condensers both in flat heat-pipes and in hybrid vapor chambers [6, 10] for efficient cooling of power electronics. While VCs are not new, only recent investigations by Damoulakis and Megaridis [5] have successfully developed *completely* wickless vapor chambers that incorporate wettability patterning to selectively render hydrophilic pathways on a hydrophobic surface. The developed VC works without the drawbacks of capillary pressure drops in wicks that can lead to dry-out scenarios in high flux applications, while allowing for VCs to become virtually as thin as possible. Such wettability patterns have been known to improve condensation performance [11] and can also be used for high-rate pumpless transport of fluids on open surfaces [12, 13].

While numerous configurations and modifications of VCs have been studied for cooling of power electronics, we demonstrate a working, fully-wickless VC for cooling of high frequency switching power electronics. This extends earlier works done by Damoulakis [5], and shows the viability of use of wickless vapor chambers in real applications requiring cool-down of high frequency power electronics. Herein, air cooling is implemented to remove heat from the heat sink and maintain the MOSFETs below their critical failing temperature.

## II. MATERIALS AND METHODOLOGY

### A. Vapor Chamber fabrication

The developed vapor chamber consisted of an evaporator side and a condenser side. As shown in Figure 1, both the evaporator (Figure 1 – A) and condenser (Figure 1 – B) of the vapor chamber comprised of rectangular 2 mm thick copper plates (Cu-101, thermal conductivity 390.8 W / m K, McMaster-Carr) with dimensions 11.43 cm × 6.35 cm. The plates were milled to a depth of 1 mm from the sides to maintain a centered elevated area of 10 cm × 5 cm. The gasket (Figure 1 – C) (Viton Fluoroelastomer, with a durometer hardness of 75A, McMaster-Carr) was cut in dimensions matching those of the milled-out plate outline and had a thickness of 3mm.

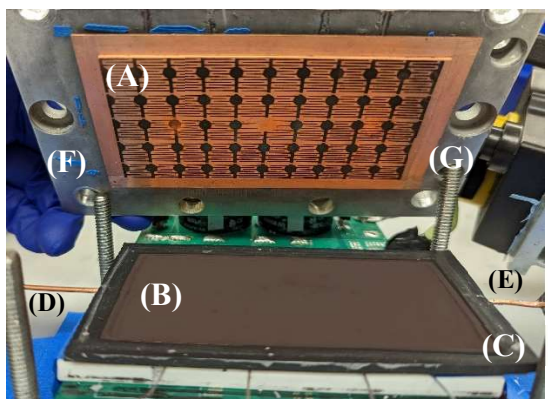

Fig. 1. Parts that make up the wick-free vapor chamber used here for cooling of the power electronics setup (not visible in this photo). The condenser plate (A) sits on top of the the evaporator plate (B), while a gasket (C) maintains proper sealing of the VC's interior from the atmosphere. (D) Vacuum port used to rid the VC of NCGs before experimentation, while a charging port (E) is used for charging the emptied VC with the working fluid (here water). Finally, the whole setup is clamped to the test bench by a 1 cm-thick aluminum plate (F) using four threaded posts (G) at its corners.

The gasket had two ports, that allowed for charging of the working fluid and for removing NCGs from the VC before experimentation. The gasket maintained the hermetic seal of the vapor chamber throughout the duration of each experiment.

Both the evaporator and the condenser plates were thoroughly cleaned via ultrasonication first in water (deionized) and then in ethanol (>98% purity) each for 10 minutes. Specific procedures were followed (below) to create wettability patterned surfaces on the condenser and the evaporator.

*Condenser*: After cleaning the copper plates, the plates were spin-coated with Teflon AF (AF-2400, Amorphous Fluoroplastics Solution, Chemours Co.). The surface was then cured in a single-zone tube furnace (Lindberg, Blue-M-HTF55322c) in a reducing environment consisting of 10% Hydrogen and 90% Argon, in three stages viz., 80 °C, 180 °C and 260 °C, each for 10 minutes, to completely cure the surface. The resulting surface was uniformly hydrophobic with static contact angle of 118° ± 2° (see Figure 2 – B) when measured with an optical goniometer. To impart hydrophilic pathways on the hydrophobic condenser, a laser-marking system from TYKMA Electrox – EMS 400, was used with a 40% power output, and a 20kHz intensity with a traverse speed of 200mm/s to etch away the hydrophobic regions (Teflon-coated areas) into desired patterns. An example of a patterned condenser is highlighted in Figure 2: the shiny parts are Teflon-coated copper and are hydrophobic, while the black laser-etched parts are hydrophilic.

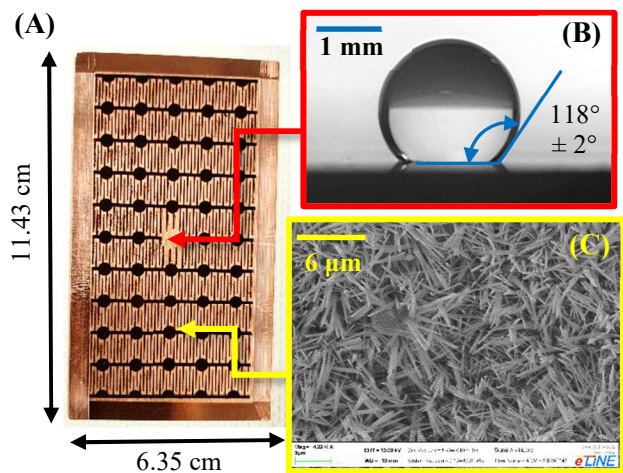

Fig. 2. (A) Patterned condenser used for experimentation in the wick-free vapor chamber. The patterns comprise of super-hydrophilic tracks (black) on a hydrophobic (copper colored) background. (B) The static contact angle of a water droplet on the hydrophobic (Teflon-coated copper) surface was estimated using optical goniometry, as shown in the top right (red inset box). The laser-etched hydrophilic pathways were subjected to chemical treatment, which grew copper hydroxide nano-hairs (C: as seen in a scanning electron microscope) thus rendering the dark pathways superhydrophilic.

After the entire pattern has been etched onto the sample, the sample was immersed into a solution of 2.5 molar NaOH (Sodium hydroxide, Sigma Aldrich – 415413 – 500mL) and 0.1 molar $(NH_4)_2S_2O_8$ (Ammonium persulfate ≥ 98%, Sigma Aldrich – MKCF3704). This chemical treatment was performed at room temperature and formed copper hydroxide nano-hairs

(as shown in Figure 2 – C) that grew on the laser-etched surface, resulting in a superhydrophilic surface [14]. Thus, the generated surface features superhydrophilic tracks on a Teflon-coated hydrophobic background.

*Evaporator*: The developed wick-free vapor chamber consisted of a uniformly superhydrophilic evaporator surface, with no patterns. The dimensions of the evaporator copper plate matched with those of the condenser; the plate was cleaned using a similar process as explained earlier, via ultrasonication in ethanol and water. The cleaned plates were then etched with a laser marking system from TYKMA Electrox – EMS 400, with a 40% power output, and a 20 kHz intensity with a traverse speed of 200 mm/s. Such laser etching imparted surface roughness to the sample, and rendered it hydrophilic. After laser etching, the sample was immediately immersed into a solution of 2.5 molar NaOH (Sodium hydroxide, Sigma Aldrich – 415413 – 500mL) and 0.1 molar $(NH_4)_2S_2O_8$ (Ammonium persulfate ≥ 98%, Sigma Aldrich – MKCF3704). This chemical treatment at room temperature formed copper hydroxide nano-hairs (as shown in Figure 2 – C) that grew on the laser-etched surface, resulting in a superhydrophilic surface with a contact angle less than 5 degrees [14].

*B. Experimental Setup*

An experimental setup (as shown schematically in Figure 3) was developed for testing the performance of the VC in cooling of high frequency switching MOSFETs. As seen in Figure 3, the vapor chamber was placed directly on top of the MOSFETs, and a 1 cm thick aluminum plate (which was air-cooled from the top and thus functioned as a heat sink) was screwed on top of the entire setup such that the vapor chamber remained in place throughout the duration of the experiment. Thermal gap pads (manufactured by Bergquist Company, with thickness of 0.25 mm and thermal conductivity of 3 W/mK) were used to maintain proper thermal contact without air gaps in between the MOSFETs and the evaporator side of the vapor chamber. A Teflon block was used to surround the MOSFETs, and ensured that heat can only escape from the MOSFETs via the VC.

Two ports located on either side of the gasket were used for degassing the system or for charging of the working fluid. These ports were connected to operating valves that could be switched off to isolate the VC from any external components once the system was ready. Six T-type thermocouples (Omega, bead diameter 0.05 mm) were placed in strategic locations (T1 – T6 in Fig. 3), and recorded temperature on both MOSFETs and the on the condenser side of the VC. The T-type thermocouples were in-turn connected to an Omega DAQ: Data Acquisition Device (USB 2400) that could record the temperature information from all thermocouples at a sampling rate of 1 Hz.

*Power Electronics assembly*: The MOSFETs implemented in heating the VC were powered and switched by a printed circuit board (PCB). The schematic of the switching circuit is provided in Figure 4 – A. An external voltage $V_{in}$ was applied to the circuitry, which switched on and off the parallelly connected MOSFETs together to produce heat to be drained by the vapor chamber. As seen from Figure 4, the setup implemented two Si-based N-channel MOSFETs of TO-247-3 packaging having 5 Ω conduction resistance, labelled $Q_1$ and $Q_2$. When both MOSFETs were switched on and off, multiple losses, viz., turn-on loss, conduction loss and turn-off loss took place in the MOSFETs, which were dissipated as heat, and were taken up by the vapor chamber. During switch-off cycles, both MOSFETs were closed, and the current passed unhindered through the diodes $D_1$. This freewheeling ensured faster MOSFET switch-off and reduced voltage stresses that may have appeared across the MOSFETs.

The external voltage $V_{in}$ was supplied from a DC power supply HY10010EX (variable output: 10 V to 60 V DC), whereas another constant power source of 20 V (Agilent E3646A) was used to power the logic circuitry and for controlling the MOSFETs. A waveform generator (Wavetek-395) sent in square waves with a frequency of 100 kHz, and controlled the switching frequency of the two MOSFETs. A digital oscilloscope (Tektronix-TPS2024B) was used to verify the signals generated from the waveform generator and monitor pertinent electrical signals.

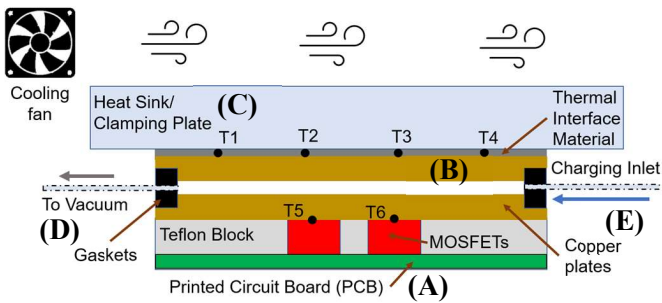

Fig. 3. Power electronics setup assembled with the present vapor chamber assembly. The MOSFETs (A) on the power electronics board switched at high frequencies, and the generated heat was subsequently transferred to the vapor chamber assembly (B). (C) An aluminum plate held the setup together during experimentation, while a constant-speed fan removed heat from the top of the plate. Two ports on either side of the VC, made way for vacumming (D) or charging (E) the setup with the working fluid (water).

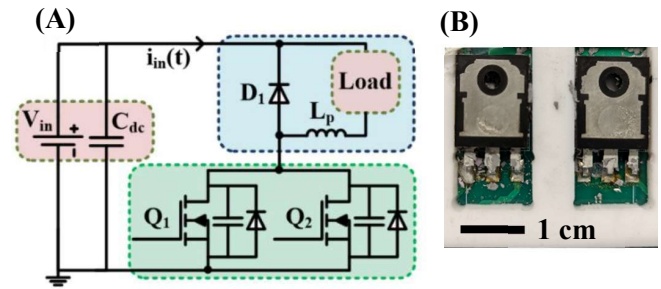

Fig. 4. (A) Schematic of the power electronics circuit that switched the MOSFETs ($Q_1$ and $Q_2$) for generating heat. (B) The actual TO-247-3 packaging MOSFETs ($Q_1$ and $Q_2$) from the circuit that interacted with the wick-free VC. The MOSFETs have a total area of 3.11 $cm^2$ that can transfer heat directly to the overlaying vapor chamber.

*Experimental procedure*: After all the components were prepared, the vapor chamber was assembled on top of the MOSFETs, and sealed with the gasket. Thereafter, a vacuum pump (Alcatel Annecy – 2008A) connected to the vacuum port was switched on, until a pressure of ~6.5 kPa was reached, and almost all NCGs from the VC were removed. The vapor chamber was left for more than an hour in its evacuated state, to ensure that the setup could hold the vacuum without any leaks, for the entire duration of the experiment. Thereafter, the VC was charged with a pre-determined amount of the working fluid, and all ports connecting the vapor chamber via the gasket were closed. The power electronics setup was turned on and the experiments were carried out. Each experimental trial at a particular heat input was carried out for 5 minutes, and the temperature data from the last 60 seconds (corresponding to a 5 minute run at a particular heat input) were used to calculate the relevant parameters. For each set of experiments, the power was continuously increased until the temperature on the MOSFETs reach 100 °C, beyond which the MOSFETs stopped working properly, or until the system reached a thermal dry-out (as indicated by a sharp increase in the readings of the evaporator thermocouples).

*C. Performance Metrics and Data Reduction*

As explained earlier, the performance of each VC at a particular power input and operating temperature was evaluated by comparison of the total thermal resistances. As shown in Eq. 1, the overall thermal resistance is defined as the ratio of the difference in temperatures between the evaporator and condenser side of the VC to the power input to the VC

$$R_{total} = \frac{\Delta T}{Q} = \frac{T_{evap}^{avg} - T_{cond}^{avg}}{Q_{in}} \quad (1)$$

The performance of the VC also depends on the quantity of its working fluid, which is designated by its charging ratio. Charging ratio, as shown in Eq. 2, is defined as the ratio of the volume of the working fluid to the entire vapor space volume

$$CR = \frac{Volume\ of\ working\ fluid}{Vapor\ Space} \quad (2)$$

Each experimental trial on the vapor chamber was conducted on a fixed *CR* and rising heat loads until the VC reached dry-out conditions. After successful completion, the VC was allowed to reach room temperature, after which, more fluid was introduced in the chamber, and subsequent experimentation was carried out for the new *CR* value. A Gaussian error propagation analysis [10, 6] was carried out to estimate experimental errors for different calculated parameters.

### III. RESULTS AND DISCUSSION

To evaluate the performance of the VCs in high frequency switching power-electronics cooling, two different testing scenarios were implemented. First, experimental trials were conducted on a wick-free VC, with a uniformly hydrophobic condenser and uniformly superhydrophilic evaporator. A Teflon-coated machined copper plate, with no patterns, acted as a uniformly hydrophobic condenser, with a water static contact angle of 118° ± 2°. Under pure steam conditions, as is prevalent inside the VC, hydrophobic surfaces have been reported to have a heat transfer coefficient an order of magnitude higher than hydrophilic cases [11]. For the evaporator, a uniform superhydrophilic surface covered with copper hydroxide nano-hairs was used. This case with a uniformly superhydrophilic evaporator and a hydrophobic condenser was used as a benchmark for testing and evaluation of the wick-free VCs.

Experimental trials were thereafter carried out on the wick-free control VC for different working fluid charging ratios. The variations in the thermal resistances are plotted in Figure 5.

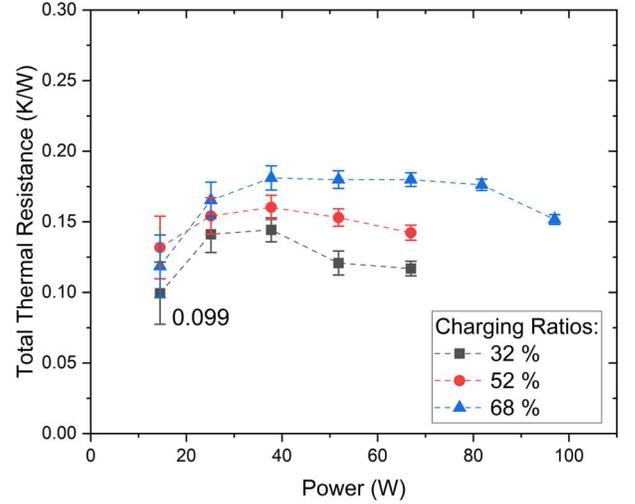

Fig. 5. Total thermal resistance vs. power input to the wick-free control VC (comprising of a uniformly hydrophobic condenser and a uniformly superhydrophilic evaporator) at varying fluid charging ratios. The lowest charging ratio of 32% was seen to perform the best, with lowest thermal resistance of 0.099 K/W recorded at power input of 14.5 W.

The thermal resistance was seen to first rise and then decrease for each *CR* with increasing power input to the VC. A minimal total thermal resistance (0.099 K/W) was observed when the VC had a low *CR* of 32 % and a power input of 14.5 W. Increasing the volume of working fluid had a negative influence on the total thermal resistance. While greater *CR*s generally do not lead to dry-out and can work even at higher heat inputs, while maintaining optimal MOSFET temperatures, they do suffer from greater thermal resistances, which can be attributed to the increased thickness of the working fluid layer present on the evaporator side, which raises thermal resistance. This hypothesis explains the trend of the increasing thermal resistance for greater *CR*s in Fig. 5.

Similar experiments were carried out on another wick-free VC featuring a patterned condenser and a uniformly superhydrophilic evaporator. Thermal resistances were also calculated for these set of experiments and has been plotted for varying power inputs and charging ratios in Figure 6.

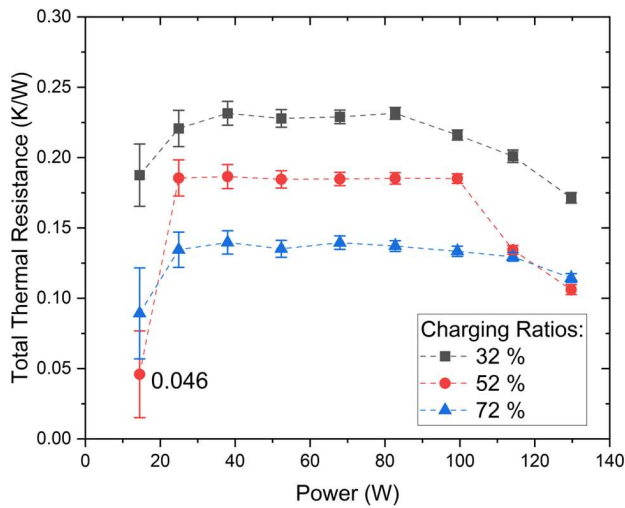

Fig. 6. Total thermal resistance vs. power input to a wick-free VC (comprising of a wettability-patterned hydrophobic condenser and a uniformly superhydrophilic evaporator) at varying charging ratios. The highest charging ratio of 72% was seen to perform the best overall, while the lowest thermal resistance of 0.046 K/W corresponded to a power input of 14.5 W and a *CR* of 52%.

As seen from Figure 6, the thermal resistance at each *CR* behaved like the benchmark cases in Fig. 5, first increasing, then leveling off and then decreasing at higher power inputs. A minimal total thermal resistance (0.046 K/W) was observed for a *CR* of 52 %, and at power input of 14.5 W. Irrespective of the *CR*, all the systems were seen to function at optimal MOSFET temperature and without thermal dry-outs. Unlike the benchmarking VC, herein the total thermal resistance was seen to decrease with increasing *CR*. Patterning the condenser resulted in uniform condensation throughout the span of the condenser surface. The patterns limited the maximum droplet dimension, while collecting the droplets in the circular wells designed for draining the working fluid to the evaporator [5]. The superhydrophilic wells stored water in the form of droplets, and drained them into the evaporator side when the height of the accumulated condensate volume exceeded the vapor space width (i.e., 1 mm, for the present VC). As explained earlier, higher *CR* not only increased the thermal resistance by adding onto conduction resistance of the increased fluid layer, but also cut down on the vapor space, thereby limiting the maximum dimension of a droplet that could exist on the condenser. Thus, the increasing fluid layer on the evaporator decreased the maximum allowable droplet size on the condenser; thus, bigger droplets could no longer exist on the condenser and had to drain (or coalesce) into the evaporator pool. This opened newer sites for nucleation, and hence enhanced condensation, thereby decreasing the overall thermal resistance of the system.

It is important to note here that while the thermal resistance values obtained in the case of a VC with a non-patterned condenser and a CR of 35 % (Figure 5) are lower than the best performing patterned VC at around 60 W, this case cannot be considered as the best-case scenario. This is because lower CRs are vulnerable to thermal dry-outs and can hamper the performance of the system. The patterned VC is seen to maintain a low and almost constant thermal resistance for a wider range of thermal inputs and clearly outperforms the unpatterned system.

## IV. Conclusions

A wick-free vapor chamber was developed for forced air cooling of high frequency switching power electronics. The vapor chamber had overall thickness of 5 mm, with a 1 mm vapor space. Distilled water was used as the charging fluid, which evaporated from a laser-etched evaporator, and condensed on a hydrophobic or patterned condenser. The water then flowed back to the evaporator, and this process continued in cycles to take away heat from two MOSFETs powered by a PCB. The MOSFETs had a TO-247-3 packaging and a high conduction resistance, and were used to simulate typical heat loads at semiconductor junctions, as observed in high frequency switching of power electronic systems.

Two distinct vapor chambers were used to maintain the MOSFETs, below a critical temperature of 100 °C beyond which the MOSFETs would fail. As a benchmark for comparison, a wickless vapor chamber comprising of a uniformly hydrophobic condenser (Teflon-coated copper), and a uniformly superhydrophilic evaporator (copper hydroxide nano hair arrays grown on laser etched copper) was used for cooling the same power electronics. Thereafter, a bio-inspired wettability pattern, which is commonly used for betterment of condensation heat transfer [11, 13], was used as the VC condenser, along with a uniformly superhydrophilic evaporator similar to the one used in the benchmarking scenario. Thermal resistances for varying charging ratios and power inputs to the VC were estimated and compared. Earlier experiments by Jafari Gukeh et al. [6] have compared the effectiveness of their patterned VC against a solid copper plate with similar dimensions. While such comparisons can elucidate the effectivity of the VC and demonstrate its capabilities in thermal management, they have not been carried out here, as more emphasis has been given to understand the effectiveness of the wettability patterns in wick-free VCs.

It is interesting to note that the patterned condensers play a significant role in altering the inner workings of the VC, and hence, its thermal resistance. As seen in the case of a wick-free VC with a uniformly hydrophobic condenser, there does not exist significant difference in thermal resistances for varying charging ratios. We hypothesize that localized condensation takes place in such scenarios, and the entire condenser fails to perform optimally, to provide better results. However, in the case of a wick-free VC with a patterned condenser, significant improvement of thermal resistance was observed at higher *CR*. The VC with the patterned condenser clearly outperformed the VC without a patterned condenser. It appears that the patterned condenser allows the VC to perform even at much greater power inputs, without dry-out, irrespective of *CR* changes.

Altogether, we demonstrated that wick-free vapor chambers are well suited for cooling of power electronics via forced air cooling. Future directions involve reducing the vapor chamber thickness, and hence moving towards development of much thinner wick-free vapor chambers.

ACKNOWLEDGMENT

This material is based upon research supported by, or in part by , the US Office of Naval Research under award number N14-20-1-2025 to the University of Nebraska, Lincoln and a subaward to the University of Illinois Chicago. The authors thank J. Rodriguez (Assistant Director) and T. Bruzan (Laboratory Mechanic) at the UIC College of Engineering Scientific Instrument/Machine shop for machining the samples. The authors would also like to thank the UIC College of Engineering, Nanotechnology Core Facility (NCF), for further development and characterization of the samples used in this work.